\documentclass[fleqn,usenatbib]{mnras}

\usepackage{graphicx, natbib, enumerate, txfonts, color, bm}

\def\mum{$\mathrm{\mu}$m}

\def\densitycgs{$\mathrm{g\,cm^ {-3}}$}
\def\dialarge{47.60}
\def\diasmall{25.14}

\title{Simulation and experiment of gas diffusion in a granular bed}

\author[C. G{\"u}ttler et al.]{C. G{\"u}ttler$^{1}$,
	      M. Rose$^{2}$,
				H. Sierks$^{1}$,
        W. Macher$^{3}$,
        S. Zivithal$^{3}$,
        J. Blum$^{4}$,
        S. Laddha$^{3}$,
        B. Gundlach$^{5}$,
        G. Kargl$^{3}$
	\\ \\
	$^{1}$ Max Planck Institute for Solar System Research, Justus-von-Liebig-Weg 3, 37077 G{\"o}ttingen, Germany\\
	$^{2}$ Ingenieurbüro Dr.-Ing. Martin Rose, Sommerhofenstra{\ss}e 148, 71067 Sindelfingen, Germany\\
	$^{3}$ Space Research Institute, Austrian Academy of Sciences, Schmiedlstra{\ss}e 6, 8042 Graz, Austria\\
	$^{4}$ Institut f\"ur Geophysik und extraterrestrische Physik, Technische Universit\"at Braunschweig, Mendelssohnstra{\ss}e 3, 38106 Braunschweig, Germany\\
  $^{5}$ Institut für Planetologie, Westfälische Wilhelms-Universität Münster, Wilhelm-Klemm-Str. 10, 48149 Münster, Germany		
}

\date{Accepted XXX. Received YYY; in original form ZZZ}
\pubyear{2023}

\begin{document}
\label{firstpage}
\pagerange{\pageref{firstpage}--\pageref{lastpage}}
\maketitle

\begin{abstract}
The diffusion of gas through porous material is important to understand the physical processes underlying cometary activity.
We study the diffusion of a rarefied gas (Knudsen regime) through a packed bed of monodisperse spheres via experiments and numerical modelling, providing an absolute value of the diffusion coefficient and compare it to published analytical models.
The experiments are designed to be directly comparable to numerical simulations, by using precision steel beads, simple geometries, and a trade-off of the sample size between small boundary effects and efficient computation.
For direct comparison, the diffusion coefficient is determined in Direct Simulation Monte Carlo (DSMC) simulations, yielding a good match with experiments.
This model is further-on used on a microscopic scale, which cannot be studied in experiments, to determine the mean path of gas molecules and its distribution, and compare it against an analytical model.
Scaling with sample properties (particle size, porosity) and gas properties (molecular mass, temperature) is consistent with analytical models.
As predicted by these, results are very sensitive on sample porosity and we find that a tortuosity $q(\varepsilon)$ depending linearly on the porosity $\varepsilon$ can well reconcile the analytical model with experiments and simulations.
Mean paths of molecules are close to those described in the literature, but their distribution deviates from the expectation for small path lengths.
The provided diffusion coefficients and scaling laws are directly applicable to thermophysical models of idealised cometary material.
\end{abstract}


\section{Introduction}

The explanation of cometary activity on the microscopic level remains one of the major challenges in cometary science with implications for the formation of small bodies (and planetesimals) in the solar system.
The sublimation of subsurface ices creates gas, which flows through the porous surface layers and creates a pressure that must be sufficient to overcome cohesive forces of the dust and lift it \citep[e.g.,][]{GundlachEtAl:2020, FulleEtal:2020}.
The pressures usually remain low enough such that the diffusion takes place in the Knudsen regime.
Many aspects on cometary near-surface layers were studied in computer simulations by \citet{SkorovEtal:2011, SkorovEtal:2021, SkorovEtal:2022}, which include packings of varying porosity, porosity inhomogeneities, or hierarchical structures.

Experimental work on the rarefied gas flow through granular beds, aiming to represent cometary surface layers, was recently performed by \citet{SchweighartEtal:2021}, supported by numerical simulations by \citet{LaddhaEtal:2023}.
That work is focused on the transition regime between viscous and Knudsen flow.
Working with realistic astrophysical samples (i.e., lunar and asteroid analogue material), the work showed the challenges of this multi-parameter problem including particle shape (angularity, sphericity) or friction, which control microscopic parameters like pore size and shape as well as macroscopic parameters like porosity.
Opening of macroscopic voids in the samples were observed, which can affect the measurements and should be avoided.
Earlier measurements by \citet{GundlachEtal:2011b}, which were focused on the sublimation of ice but contain measurements of the diffusion coefficient appear non comparable to the measurements by \citet{SchweighartEtal:2021}, which was our motivation for a fundamental experiment.

In this work, we focus on the Knudsen regime and the measurement of the Knudsen diffusion coefficient for a wide range of parameters.
\citet{Knudsen:1909} provided a general description of this diffusion process and the coefficient itself, which is used by many authors \citep{SkorovEtal:2011, GundlachEtal:2011b, SchweighartEtal:2021}, but strictly applicable only to a geometry that can be approximated by flow tubes.
A mathematical statistical model by \citet{Derjaguin:1946} is independent of geometry and supported by laboratory measurements and independent derivations by \citet{AsaedaEtal:1973}.
This model will be reviewed in Sect. \ref{sect:theoretic_background} and applied to our geometry of packed beds of monodisperse spheres.

We then provide experiments and numerical simulations to determine the diffusion coefficient.
Our samples are on the one hand idealised to allow a match between experiments and simulations.
On the other hand, we expect that a primordial cometary surface, formed from dust pebbles under low gravity \citep{BlumEtal:2014} is reasonably represented by a packing of spheres.
With our numerical simulations, we thus intend to look into the microscopic detail of gas diffusion for realistic packings.
In a first step, provided in this article, we want to establish a robust link between the numerical simulations, experiments, and analytic models.
This will in the future allow an extrapolation of the model to more complex geometries, while maintaining an understanding of the link between microscopic effects and macroscopic parameters (i.e., diffusion coefficient).
Applications could be porosity gradients, polydisperse and irregular particles, macroscopic cracks, gas production inside the sample, and many others.
Moreover, the model is capable of studying the micro-physics of the liftoff process and further outgassing properties \citep{ChristouEtal:2020}, and can be used for the interpretation of complex cometary physics experiments \citep[see, e.g.,][]{KreuzigEtal:2021}.

The theoretical background is laid out in Sect. \ref{sect:theoretic_background}, followed by our experiments described in Sect. \ref{sect:experimental_setup}.
Using the numerical samples described in Sect. \ref{sect:numerical_samples}, our Discrete Simulation Monte Carlo (DSMC) model is used in comparison to the experiments as well as analytical models and their assumptions in Sect. \ref{sect:dsmc}.
The results are compiled in Sect. \ref{sect:discussion} and we conclude the main findings of this work in Sect. \ref{sect:conclusion}.

\section{Theoretical Background}  \label{sect:theoretic_background}

In this section we lay out the required equation system for the flow of gas through a granular bed made up of monodisperse spheres.
In the free molecular flow regime (or Knudsen regime; when Knudsen number $\mathrm{Kn} \gg 1$) it is leaned on the description of \citet{Derjaguin:1946}.
It is applicable when collisions between molecules and the constraining environment (walls and inner surfaces) are dominant over collisions among gas molecules.
The relevant representative physical length scale in our case is the pore size.
The contribution of the viscous flow in our experimental work is small but non-negligible, thus needs to be considered and is also described below.

We first focus on the Knudsen regime, where the diffusive molar flow per area $\bm{j}$ [$\mathrm{mol}\,\mathrm{m}^{-2}\,\mathrm{s}^{-1}$] for a single gas species can be formulated as
\begin{equation}
    \bm{j_\mathrm{d}} = - D_\mathrm{K} \cdot \nabla n \; , \label{eq:diffusive_flow}
\end{equation}
where $n$ [mol m$^{-3}$] is the gas number density and $D_\mathrm{K}$ [$\mathrm{m}^2\;\mathrm{s}^{-1}$] is the (Knudsen) diffusion coefficient.

The diffusion coefficient is determined by the structure of the granular material through which the gas flows, as well as the gas kinetic properties.
For a porous medium, it is derived by \citet[][his Eq. 10']{Derjaguin:1946} as
\begin{equation}
        D_\mathrm{K} = \frac{1}{6} \cdot \varepsilon \cdot \overline{c} \cdot \frac{\overline{\lambda^2} - \frac{8}{13}\cdot \overline{\lambda}^{\;2}}{\overline{\lambda}}\; , \label{eq:dk_derjaguin1}
\end{equation}
where $\overline{\lambda}$ is the mean path (i.e., average distance between surface collisions of molecules, not to be mistaken with the mean free path in the gas phase) and $\overline{\lambda^2}$ is the average of their squares.
The mean gas velocity $\overline{c}$ [$\mathrm{m\;s^{-1}}$] is given as
\begin{equation}
   \overline{c} = \sqrt{\frac{8RT}{\pi M}} \; , \label{eq:gas_velocity}
\end{equation}
where $R$ [$\mathrm{J}\;\mathrm{K}^{-1}\;\mathrm{mol}^{-1}$] is the gas constant, $T$ [K] the gas temperature and $M$ [$\mathrm{kg}\;\mathrm{mol}^{-1}$] the molecular mass.
Equation \ref{eq:dk_derjaguin1} differs from the representation of \citeauthor{Derjaguin:1946} by a factor $\varepsilon$ because of the different definition of concentration wrt. our definition in Eq. \ref{eq:diffusive_flow} as also noted by \citet{AsaedaEtal:1973}.

Further on, \citeauthor{Derjaguin:1946} makes the assumption that the probability density function of the molecules' normalised path segments $\lambda' = \lambda / \overline{\lambda}$ follows an exponential distribution (defining a Poisson process) of the form
\begin{equation}
    f\left(\lambda'\right) \mathrm{d} \lambda' = e^{-\lambda'} \mathrm{d} \lambda' \; ,  \label{eq:poisson}
\end{equation}
thus that the ratio between the square-weighted mean path and the squared mean path is
\begin{equation}
    \overline{\lambda^2} / \overline{\lambda}^{\;2} = 2 \; .  \label{eq:lambdas_ratio}
\end{equation}
He provides the mean path length as $\overline{\lambda} = 4/s$, where $s$ is the specific surface.
In our geometry, we take this as the surface area of pores (equal to that of our spheres) normalised to the volume of pores and get
\begin{equation}
    \overline{\lambda} = \frac{2}{3} \cdot d_\mathrm{s} \cdot \frac{\varepsilon}{1-\varepsilon} \; , \label{eq:lambda_derjaguin}
\end{equation}
where $d_\mathrm{s}$ [m] is the sphere diameter, Eq. \ref{eq:dk_derjaguin1} can thus be written as
\begin{equation}
    D_\mathrm{K} = \frac{2}{13} \cdot d_\mathrm{s} \cdot \frac{\varepsilon^2}{1-\varepsilon} \cdot \overline{c} \label{eq:dk_derjaguin2}
\end{equation}
\citep[][his Eq. 15']{Derjaguin:1946}.
This is consistent with the description of \citet[][their Eq. 25]{AsaedaEtal:1973}
\begin{equation}
    D_\mathrm{K} = \frac{1}{3 \cdot q \cdot \Phi} \cdot d_\mathrm{s} \cdot \frac{\varepsilon^2}{1-\varepsilon} \cdot \overline{c}, \label{eq:dk_asaeda}
\end{equation}
who determined $\Phi \approx 2.18$ and measured $q=1.41$.
We will stick to $\Phi=13/6$, such that $q=1$ becomes identical to Eq. \ref{eq:dk_derjaguin2}, but keep in mind that \citeauthor{AsaedaEtal:1973} determined $q=1.41$ from experiments (further discussion of $q$ in Sect. \ref{sect:dsmc}).

If collisions between gas molecules become dominant over collisions with the medium, i.e., $\mathrm{Kn} \ll 1$, the gas is in the viscous-flow regime.
If we take the gas pressure as $p$ [Pa], the gas dynamic viscosity as $\mu$ [$\mathrm{kg}\,\mathrm{m}^{-1}\,\mathrm{s}^{-1}$], and the gas permeability as $B$ [$\mathrm{m}^2$], the molar flow in this regime follows as \citep[e.g.,][their Eq. 2]{KastHohentanner:2000}
\begin{equation}
    \bm{j_\mathrm{v}} = - \frac{p \cdot B}{\mu} \cdot \nabla n \; . \label{eq:viscous_flow}
\end{equation}
This flow becomes increasingly important as the pressure increases and consequently collisions among gas molecules become dominant.

Following \citet[][their Eq. 28]{MasonEtal:1967}, we consider the two flow contributions from Eqs. \ref{eq:diffusive_flow} and \ref{eq:viscous_flow} as independent and write the total flow as
\begin{equation}
    \bm{j} = \bm{j_\mathrm{d}} + \bm{j_\mathrm{v}} = - \left( D_\mathrm{K} + \frac{p \cdot B}{\mu} \right) \cdot \nabla n \; .
    \label{eq:linear_added_flow}
\end{equation}

In the experiments presented below, we will determine the flux $j = \bm{j}_z$ through a cylindrical sample of height $h$ along the symmetry axis in $-z$.
With the use of the ideal gas law, the average gradient of the number density is
\begin{equation}
    \overline{n'} = \overline{\partial_z n} 
    = \frac{1}{RT} \overline{\partial_z p}
    = \frac{1}{RT} \frac{p_\mathrm{u}-p_\mathrm{d}}{h} \; ,
\end{equation}
where $p_\mathrm{u}=p(z=h)$ and $p_\mathrm{d}=p(z=0)$ are the upstream and downstream pressure, respectively.
For later use, we integrate Eq. \ref{eq:linear_added_flow} to obtain an expression of the coefficients $D_\mathrm{K}$ and $B$ in terms of the measured gas flux $j$ and pressure at the sample in- and outflow sides,
\begin{equation}
    \frac{-j}{\overline{n'}} = \frac{j \cdot h \cdot R \cdot T}{p_\mathrm{u}-p_\mathrm{d}} 
    = D_\mathrm{K} + \frac{\bar{p} \cdot B}{\mu} \; , \label{eq:linear_diffusion}
\end{equation}
where $\bar{p}=(p_\mathrm{u}+p_\mathrm{d})/2$ is the average pressure inside the sample.

\section{Experimental Setup and Procedures} \label{sect:experimental_setup}

The aim is to provide a reference measurement for the flow through a granular medium of spheres.
Our focus is on the clarity of the design, to allow the simulation as described in Sect. \ref{sect:dsmc} with as few assumptions as possible.
In this section we will describe the experimental setup (Sect. \ref{sect:setup}), the selection and preparation of the sample (Sect. \ref{sect:sample}), and the experimental procedures and data reduction (Sect. \ref{sect:procedures}).

\subsection{Experimental Setup} \label{sect:setup}

\begin{figure}
  \centering
  \includegraphics[height=7cm]{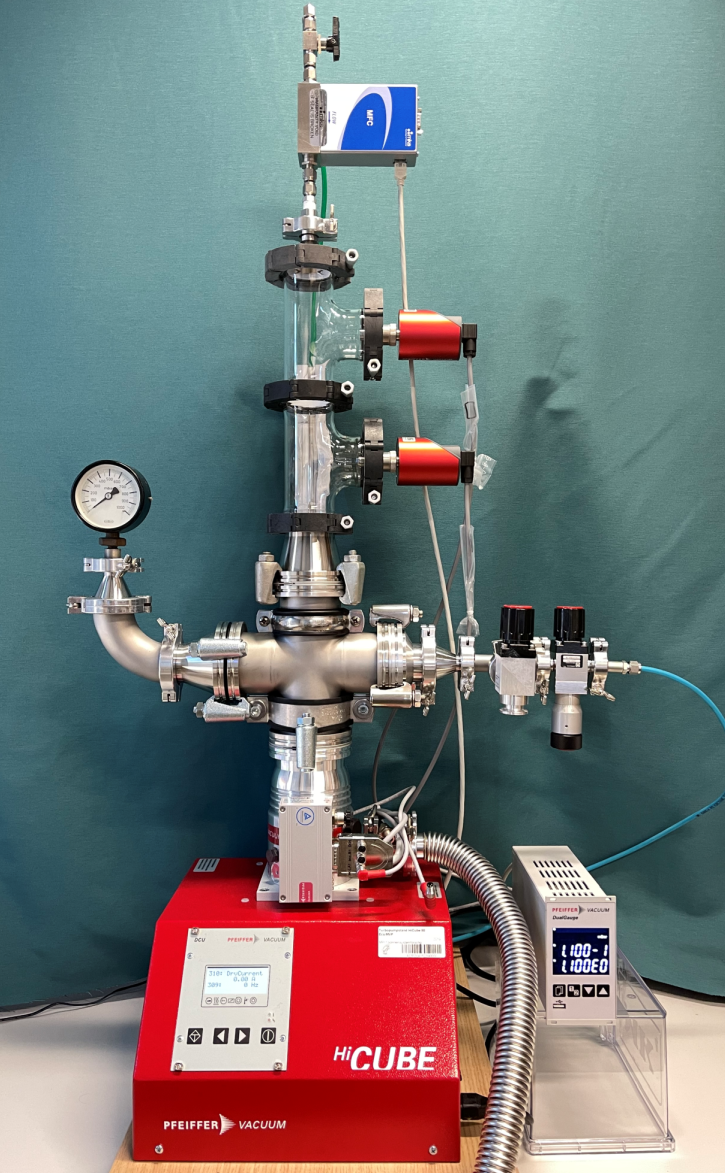}
  \hspace{0.05cm}
  \includegraphics[height=7cm]{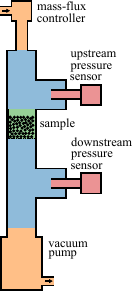}
  \caption{\label{fig:vacuum_setup}Photograph and sketch of experimental setup. The gas flows from top to bottom.}

  \includegraphics[height=2.7cm]{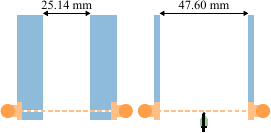}
  \hspace{0.05cm}
  \includegraphics[height=2.7cm]{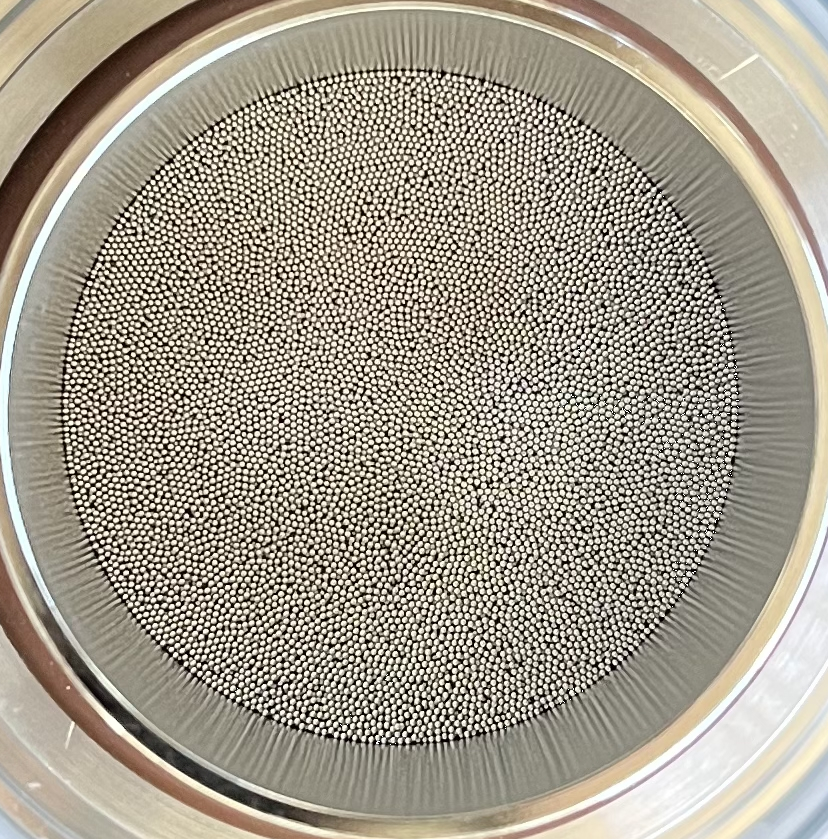}
  \caption{\label{fig:sample_container}Left: Sketch of \diasmall\ and \dialarge\ mm diameter sample containers. Both are rotational symmetric, except the green support bar below the \dialarge\ mm container. Right: Top view of the \dialarge\ mm container filled with 0.5 mm steel beads.}
\end{figure}

The experimental setup is displayed in Fig. \ref{fig:vacuum_setup}.
Following the path of the gas flow, dry $\mathrm{N_2}$ enters at the top through a mass flow controller (MKS GE50A), with selectable gas flow between 0 and \mbox{10 SCCM} (standard cubic centimetres per minute), the latter corresponding to $7.4\cdot 10^{-6}\;\mathrm{mol \; s^{-1}}$.
It fills an upstream volume, where the pressure $p_\mathrm{u}$ is measured by a pressure gauge (Pfeiffer CMR 364, 0.1 -- 110 Pa).
The gas then passes through the sample, mounted on top of an ISO-KF centre ring (the sample container will be described in detail below).
It enters into the downstream volume, where the pressure is measured with a second pressure gauge (Pfeiffer CMR 365, 0.01 -- 11 Pa), and which is pumped by a turbo-molecular-pump.

The main vacuum parts are made of glass to see the sample during the experiment and to verify that it is not modified by the gas flow (this did not turn out to be an issue).
The volume of the upstream and downstream compartments are approximately 350 and 1500 cm$^3$, respectively.
The capacitive, temperature compensated pressure sensors are mounted with Swagelok Ultra-Torr Vacuum Fittings, such that their inlet tubes reach into the glass chamber.
All surfaces were cleaned under ISO 4 cleanroom conditions, while the experiment was then operated in clean but normal laboratory conditions.
The temperature of the room was regulated and direct sunlight was avoided onto the setup before and during an experiment run.

The sample container as displayed in Fig. \ref{fig:sample_container} is based on an ISO-KF50 centre ring with a mesh in the ring plane (orange in figure).
The mesh was analysed with an optical microscope and has steel wires of 112 \mum\ diameter with a pitch of 318 \mum.
This results in an opening of 206 \mum, or an area fraction of the openings of 41\%.
The inner diameter of the centre ring is \dialarge\ mm, such that for the \dialarge\ mm container (centre), a matching steel tube (blue in figure) was welded on top of this.
Due to a bending of the steel mesh, an aluminum bar was glued below the centre ring to fix a supporting M1.6 screw in the centre of the mesh.
The bar is shown in green in the centre figure and is the only part, which is not rotational symmetric.
A photograph of the \dialarge\ mm sample container, filled with 0.5 mm steel beads is displayed on the right.
A second container with inner diameter of \diasmall\ mm is shown on the left.
Instead of the \dialarge\ mm steel tube, a form locking cylinder is welded above and below the mesh (blue).
This reduces the bending of the mesh such that it is not supported with a screw in this design.
The constant and well defined inner diameter (i.e., straight cylinder with no step) was a desired characteristic to simplify numerical modelling.
The containers were electropolished after manufacturing.

\subsection{Sample and Filling} \label{sect:sample}

As samples, we used 0.5 mm diameter precision steel beads (rolled steel 1.3505; 7.83 \densitycgs) of grade 28 according to ISO 3290.
This means, variations in the diameter or roundness of $\le 0.7$ \mum\ and surface roughness $\le 0.05$ \mum.
300,000 steel beads were procured, which result in a maximum filling of 18 mm of the \dialarge\ mm container.
To remove potential residual oils on the surface, the spheres were cleaned in an ultra-sound bath of isopropyl alcohol, then acetone, and baked out under reduced pressure.
With a UV lamp we confirmed that the sample was not contaminated with dust particles.
Some steel beads appear slightly magnetic, i.e., sticking to container walls or tweezers.
However, no effect on the collective (flow) behaviour was observed during handling, so it is not expected to affect the granular structure.

\begin{figure}
  \centering
  \includegraphics[width=\columnwidth]{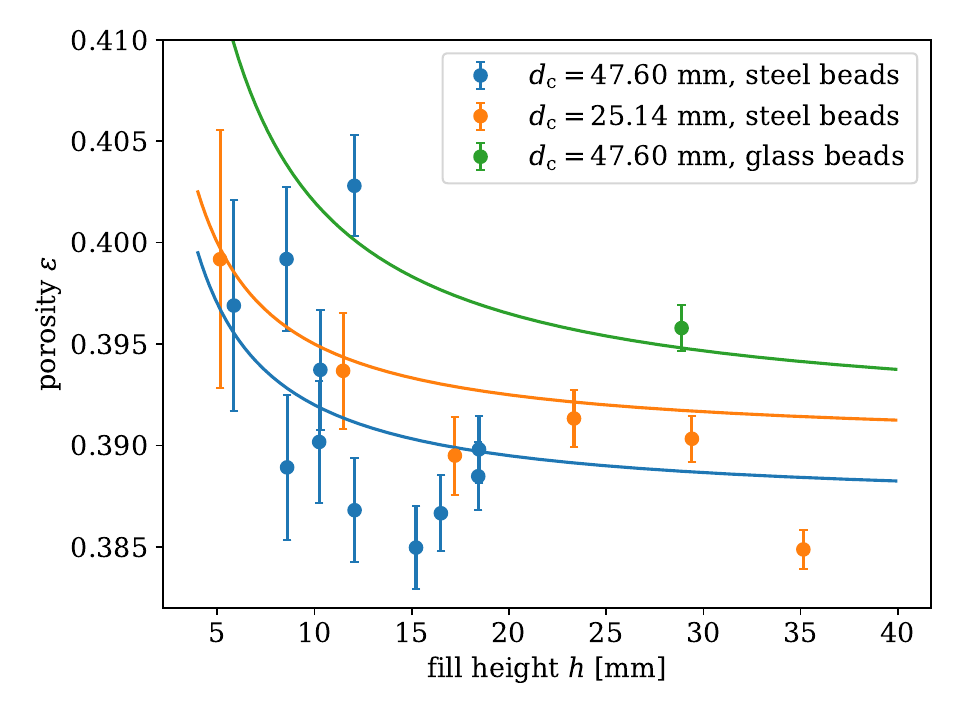}
  \caption{\label{fig:porosity_over_height}Porosity of samples. The solid lines are model approximations described in the text.}
\end{figure}

The spheres were gently filled into the sample container, which was then vertically shaken (or rolled) to even the surface.
After that, a fitting cylinder was pushed onto the upper surface (approx. 3 N) and rotated by 360$^\circ$.
Among different tested procedures, this resulted in the smallest variations in porosity.
The height of the cylinder relative to the container rim was measured to a precision of 1 \mum, but showing variations of 10 -- 20 \mum\ for individual measurements due to the sample yield.
From this measurement the fill height can be determined, so that the sample porosity can be calculated from the measured mass and the given density of the spheres.

An independent measurement with a cylindrical sample container with solid floor revealed that the bending of the sieve had an effect on the effective height ($\delta h = 0.45$ mm for $d_\mathrm{c} = $ \dialarge\ mm container; $\delta h = 0.2$ mm for $d_\mathrm{c} = $ \diasmall\ mm container), which was corrected for.
Overall, we estimated an error of the fill height of 50 \mum\ and a mass error of 50 mg.
Errors of sphere density and container diameter are relatively smaller, thus ignored.

Figure \ref{fig:porosity_over_height} shows the porosity of all experiments as a function of the fill height.
Variations in the porosity are larger than the errors and were confirmed to be resulting from statistical arrangement of the sample by filling tests with the solid-floor bottom container and reduced height and mass errors.

One experiment was performed with glass beads of 1.00 -- 1.12 mm (green in Fig. \ref{fig:porosity_over_height}), identical to the ones used by \citet[their ``Glass 7'']{SchweighartEtal:2021}.
The beads were cleaned and prepared with the same procedure as the steel beads and serve as a direct comparison to the experiments by \citeauthor{SchweighartEtal:2021}

The solid curves (per container, same coloring as legend) are based on simple model assumptions:
They take into account a layer of one sphere diameter on the top and bottom of the sample having a higher porosity due to boundary effects \citep[Sect. \ref{sect:numerical_samples}; also][their Fig. 11]{LaddhaEtal:2023}.
For a constant core porosity, this results in a fill-height dependent global porosity.
Core porosity, boundary porosity, and boundary height are chosen to realistically reproduce the data and fall in a plausible range.

\subsection{Procedures and Data Reduction} \label{sect:procedures}

After filling the sample container as described above, it was inserted into the vacuum chamber, which was evacuated for at least 12 hours.
The end pressure was at the lower limit of the pressure gauges, in the order of $10^{-2}$ Pa, and would not significantly fall for longer pumping.
Each experiment run was started with a flow of 8 SCCM and successively reduced in steps of 0.5 SCCM.
The pressure per flow step stabilised in the order of 10-60 seconds, depending on the container diameter and fill height.
After the pressure had stabilised, the level was held for at least one minute.
The upstream and downstream pressures were computer recorded with a 1 second cadence and averaged per level.
The same procedure was repeated at least once for each filling, without rearranging or even touching the sample in-between.

\begin{figure}
  \centering
  \includegraphics[width=\columnwidth]{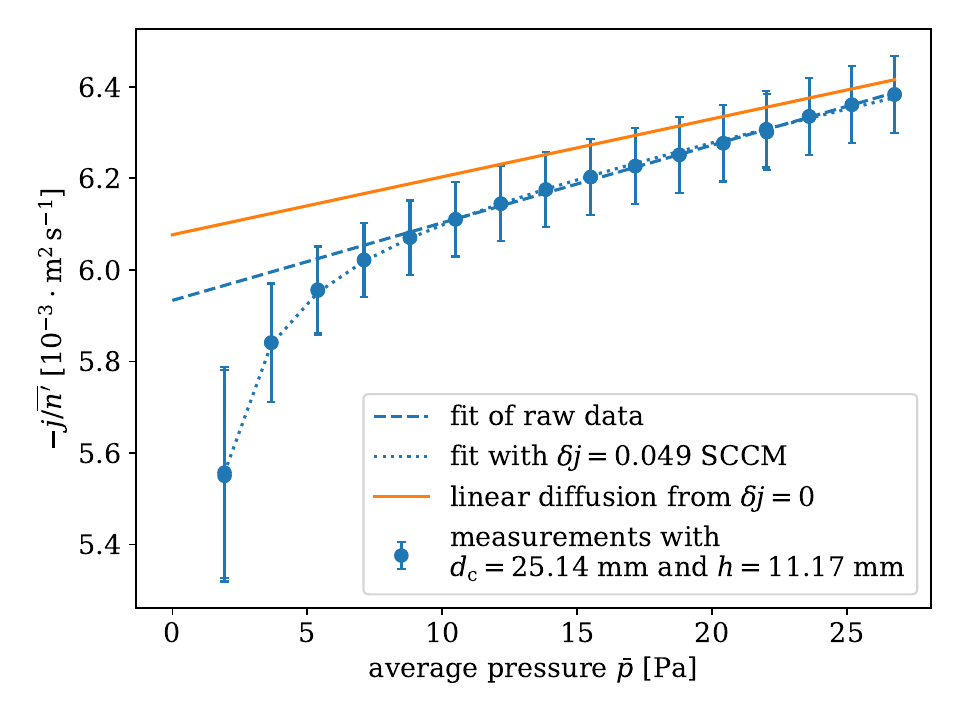}
  \caption{\label{fig:diffusion_example}Example of the diffusion measurement for one sample.}
\end{figure}

The repeatability was within the noise for the repeats, also after several days of further pumping.
An example run is shown in Fig. \ref{fig:diffusion_example} for the \diasmall\ mm container and two repeats.
The data points of the two runs are nearly on top of each other and can be distinguished only for the smallest pressure.
From Eq. \ref{eq:linear_diffusion} we would expect a linear relation between the average pressure $\bar{p}$ and the effective diffusion $-j / \overline{n'}$, which is apparently not the case, in particular for low pressures.

\begin{figure}
  \centering
  \includegraphics[width=\columnwidth]{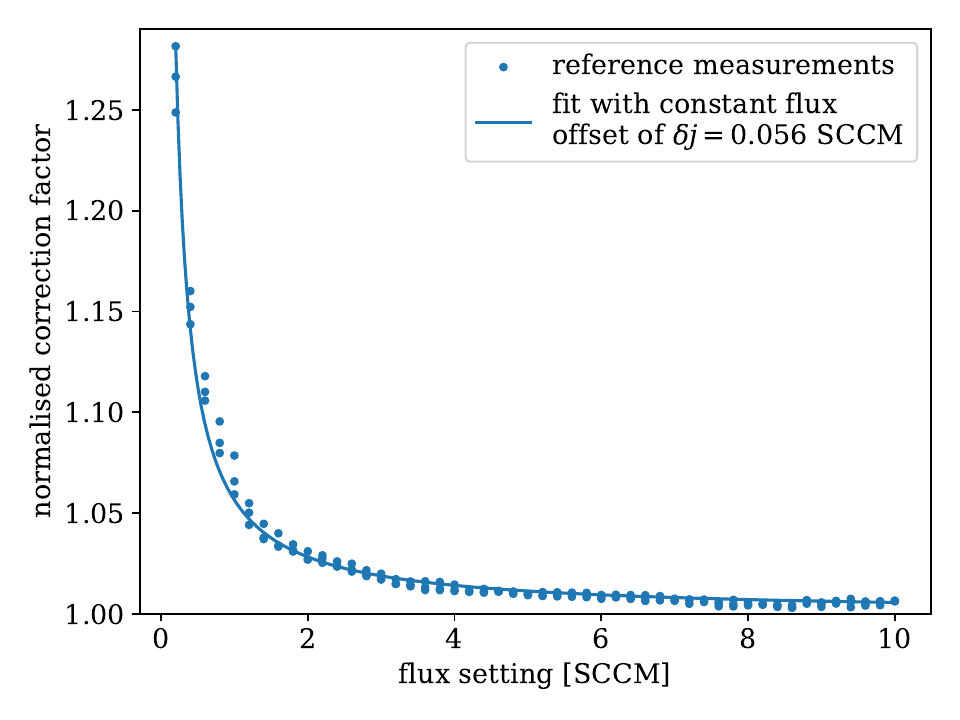}
  \caption{\label{fig:correction_curve}Flow offset of the mass-flow controller as a function of the flow setting. 
  The data can be described with a constant parasitic flow $\delta j$ as of Eq. \ref{eq:flow_correction}.}
\end{figure}

The reason for this deviation towards low pressures is a technical shortcoming of the mass-flow controller (MFC), which was therefore independently analysed.
For this purpose, the setup as of Fig. \ref{fig:vacuum_setup} was extended with a valve between the chamber volume (blue) and the vacuum pump (orange, bottom) and operated without sample container (green).
The chamber was pumped to the lowest possible pressure, then the valve was closed and the MFC set to a pre-defined flow $j_\mathrm{set}$.
The increase of the pressure was logged (on a 1 second resolution) and linearly fitted to get the pressure increase $\partial_t p$, which is proportional to the real flow $j_\mathrm{real}$.
The measurement was repeated for different flow settings from 0.5 to 10 SCCM and on three different days.

The results of this calibration experiment are shown in Fig. \ref{fig:correction_curve}, where the normalised ratio $\partial_t p / j_\mathrm{set}$ is plotted over $j_\mathrm{set}$.
For a perfect system ($\partial_t p \propto j_\mathrm{real} = j_\mathrm{set}$), this should be unity, whereas we see a strong deviation of up to 25\% for low flow settings.
The data can be well approximated assuming a constant parasitic flow contribution $\delta j$ not accounted for by the MFC (e.g., leakage through the system)
\begin{equation}
    \frac{\partial_t p}{j_\mathrm{set}} \propto \frac{j_\mathrm{set} + \delta j}{j_\mathrm{set}} \; . \label{eq:flow_correction}
\end{equation}
A similar behaviour was observed in independent experiments by \citet{ZivithalEtal:2022}.

Knowing that our measured flow must be corrected with an extra flow $\delta j$, Eq. \ref{eq:linear_diffusion} can be expanded as
\begin{equation}
    \frac{-j}{\overline{n'}} = D_\mathrm{K} + \frac{\bar{p} \cdot B}{\mu} + \frac{\delta j}{\overline{n'}} \; , \label{eq:linear_diffusion_corrected}
\end{equation}
which is then used to fit the data as in Fig. \ref{fig:diffusion_example}.
The dotted line in Fig. \ref{fig:diffusion_example} shows this fit, which well represents the data.
The parasitic flow contribution ($\delta j = 0.049$ SCCM) is not identical but very similar to the contribution determined in the calibration experiment ($\delta j = 0.056$ SCCM).
The solid curve represents Eq. \ref{eq:linear_diffusion} with the fit parameters $D_\mathrm{K}$ and $B / \mu$ from Eq. \ref{eq:linear_diffusion_corrected}.
This would be the diffusion measured with an optimal system where $\delta j = 0$.
The dashed line shows a fit only to the quasi linear part of the data to the right and it is apparent that in particular the diffusion coefficient (intercept with vertical axis) would be off by 2.4\% in this typical example.
The correction of the MFC shortcoming is thus small but non-negligible.

\begin{figure}
  \centering
  \includegraphics[width=\columnwidth]{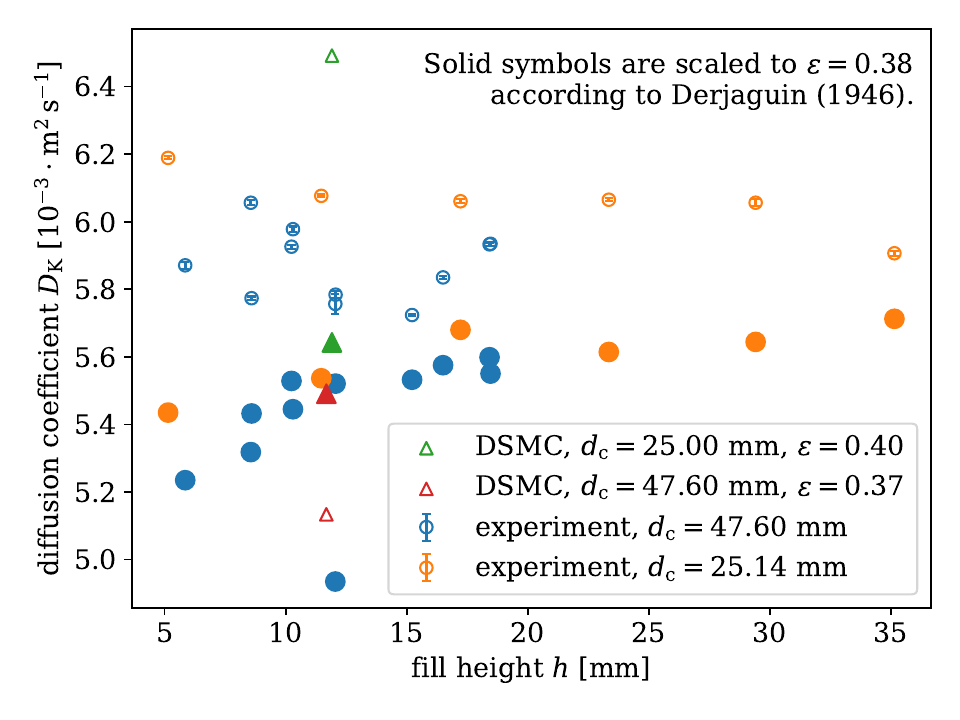}
  \caption{\label{fig:diffusion_over_height}Diffusion coefficient $D_\mathrm{K}$ for samples of different container diameter and fill height. Circles are from experiments, triangles from simulations. Open symbols are the measured values, filled symbols are scaled for porosity using $\varepsilon^2/(1-\varepsilon)$.}
\end{figure}

Measurements for different sample containers (\diasmall\ mm and \dialarge\ mm diameter) and different fill heights are shown as open circles in Fig. \ref{fig:diffusion_over_height}.
With the use of the factor $\varepsilon^2/(1-\varepsilon)$ from Eq. \ref{eq:dk_derjaguin2}, the values with measured porosity are scaled to a common porosity of $\varepsilon=0.38$ (filled circles).
For $h < 15$ mm, a slight increase over height can be observed, but the scatter is significantly reduced with respect to the open circles.
A comparison with DSMC simulations (triangles) will be drawn in Sect. \ref{sect:dsmc}.

\section{Numerical Granular Samples} \label{sect:numerical_samples}

Granular samples of monodisperse spheres were generated by two methods.
The samples to be directly comparable to the experiments from Sect. \ref{sect:experimental_setup} were created using the open source discrete element method LIGGGHTS \citep{KlossEtal:2012}.
Samples with higher porosities were created with an algorithm we call Selective Ballistic Deposition (SBD).

To construct a sample in LIGGHTS, spheres were created on a cylindrical cross section above a cylindrical volume with a solid bottom and dropped from a height of 200 sphere diameters.
The spheres interact and bounce and the simulation is run until the kinetic energy fluctuates around a small value.
Sphere interaction follows a physical model in LIGGGHTS but the parameters were not optimised for realistic material properties.
Instead, the gravity was changed per simulation, resulting in a final porosity in the desired range.
A total of 350,000 spheres were dropped, of which the lower 200,000 were used, which improves the flatness of the upper surface.

As a second method to create samples with higher porosities in the range of 0.4 to 0.85 we use Selective Ballistic Deposition \citep[SBD; similar to][]{KlarEtal:in_prep}.
The method is based on the well known Random Ballistic Deposition (RBD) method.
In the idealised RBD method, individual spherical particles are deposited unidirectionally onto a plane perpendicular to the deposition direction ($-z$ in our case).
They are deposited upon first contact with either the plane or a previously deposited particle (hit and stick).
For large samples, the porosity converges to 0.85 \citep{WatsonEtal:1997, BlumSchraepler:2004}.
The SBD method has one further constraint, namely that for each deposited particle, $\nu$ particles are test-deposited and only the particle with the smallest $z$ component is kept.
The number of test depositions $\nu$ determines the porosity of the bulk sample, which ranges from $\sim 0.4$ ($\nu \rightarrow \infty$) to 0.85 ($\nu=1$; RBD).
Details of the SBD structure in comparison to three other sample-construction algorithms are described by \citet{KlarEtal:in_prep}, who also provide many structural features and constraints on these samples (homogeneity, coordination number, contact isotropy, and others).

The idea for using the SBD method in this work is that one single algorithm can produce agglomerates of different porosities but of a similar, consistent structure.
With our implementation, we achieved a minimum porosity of 42\% with $\nu=95,961$ for an agglomerate with $320,000$ spheres and a cylindrical cross section of diameter $95.2 \cdot d_\mathrm{s}$.
This porosity is calculated in a cylindrical volume that encloses all monomers.

\begin{figure}
  \centering
  \includegraphics[width=\columnwidth]{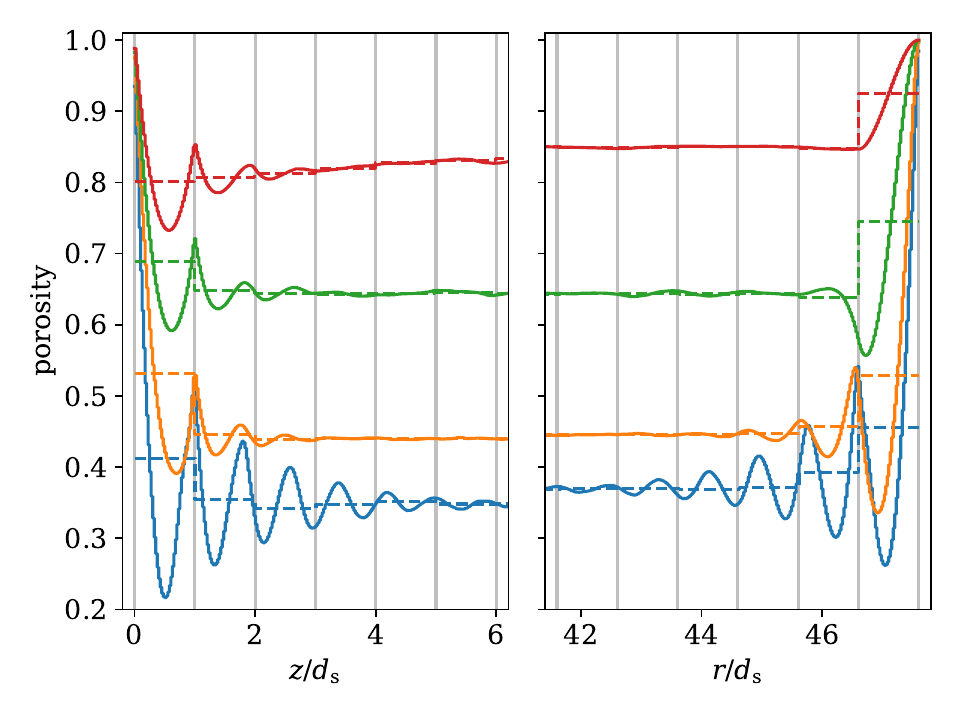}
  \caption{\label{fig:porosity_waves}Vertical (left) and radial (right) porosity variations at boundary of samples with average porosity 84\% (RBD, red), 65\% (SBD, green), 45\% (SBD, orange), and 37\% (LIGGGHTS, blue).}
\end{figure}

We analysed the porosity near the deposition plane ($z=0$) by computing average porosities in horizontal plane sections much thinner than a sphere diameter (Fig. \ref{fig:porosity_waves}, left).
The porosity is 1 at the deposition plane and undershoots below the average value at $z \approx 0.5\cdot d_\mathrm{s}$, because spheres in this layer are all at the same $z$ coordinate and have their largest extend at the sphere centre.
The porosity oscillates further, because the next layer is constrained by the bottom layer.
This 'sorting effect' and consequently the oscillations continue into the sample by a few sphere diameter (gray lines).
A similar effect is observed when computing the porosity in cylinder shells near the constraining sample-container wall (Fig. \ref{fig:porosity_waves}, right; in this case, we determine the porosity based on the elliptic integral method provided by \citet{BoersmaKamminga:1961}).
The dashed lines show the same analysis but the porosity is averaged in planes (left) or cylinder shells (right) of one sphere diameter thickness.
This is quantitatively comparable to the analysis of \citet{LaddhaEtal:2023} and expected to be the same as in experimental samples.
The consequence of these boundary effects will be further discussed in Sect. \ref{sect:boundary_effects} below.

The analysis also showed a vertical gradient in the LIGGGHTS samples (offset of blue line from left to right panel).
Due to the gravity, the porosity is lower at the bottom than at the top, both deviating 2\% in porosity from the average.
We consider this as a $\pm 2\%$ error in porosity.
The offset between the red lines from left to right panel is due to a porosity gradient in the lower 20 layers of the RBD sample resulting from the low opacity of the sample during initial deposition.
Consequently also these samples are considered with $\pm 2\%$ porosity error for the lower layers.

\section{Direct Simulation Monte Carlo} \label{sect:dsmc}

\subsection{Method}

Direct Simulation Monte Carlo (DSMC) is a method to describe the collective behaviour of gas molecules, represented by a statistically sufficient number of super molecules (in the following molecules).
The method was originally described by \citet{Bird:1994}, our implementation PI-DSMC is described in \citet{Rose:2014}.
The code passes common validation tests, i.e. mass and energy are conserved in open systems with walls, the collision rate is in accordance with theoretically predicted values and the code reproduces mass flow rates through orifices and channels as reported in the literature.
It was applied in the cometary context by several authors \citep{GicquelEtal:2017, ShiEtal:2019, ShiEtAl:in_prep}.

The method is described by \citet{Bird:1994}, thus we will provide only a short summary here.
In a volume divided into grid cells, motions and velocities of molecules are computed in an alternating, iterative way.
Time steps between these computations are chosen to be short enough such that the fastest molecule cannot cross a cell in one step.
The computation of the molecules' motion contains interactions with solid surfaces, the interaction with these is computed as diffuse reflection.
Within the cells, random pairs of molecules collide using a variable hard sphere model.
Molecules do not have any degree of freedom in rotation or vibration.
For each cell and every 10 steps, velocity components, their squares and molecule numbers are summed up (time period must be long enough to reduce noise).
The temperature follows from statistical physics, the pressure from the local density and velocity distribution of the gas molecules.
Motions and collisions are computed until the number of molecules and the gas density distribution remains constant, i.e., the simulation has reached an equilibrium state.

The output files provide the statistical parameters number density, component-resolved velocity, temperature, and pressure per grid cell.
These are sufficient to determine the diffusion coefficient.
Optionally, to study the microphysics, outputs are generated for a pre-determined number of molecules, which include molecule indices, time sorted locations and types of their collisions.
Types are either molecule-wall or molecule-molecule collisions.

We use $T=298$ K and $M=28\;\mathrm{g\;mol^{-1}}$ (N$_2$) throughout this chapter if not otherwise mentioned.

\subsection{Comparison to Experiments} \label{sect:dsmc_experiment}

The samples used to reproduce the experiments with 0.5 mm steel beads were produced with LIGGGHTS (see Sect. \ref{sect:numerical_samples}).
The sample diameter is 47.60 and 25.00 mm and the gravity was set to achieve a porosity of 37\% (47.60 mm) and 40\% (25.00 mm).
As in the experiments, the flow into an upstream compartment of 1 mm height was set to a constant value.
Molecules reaching the downstream compartment were removed from the simulation ($p_\mathrm{d} = 0$ Pa).
The simulation was run until the number of simulated molecules reached an equilibrium and the pressure $p_\mathrm{u}$  in the center plane of the upstream compartment was determined.

The resulting diffusion coefficients are plotted as open triangles in Fig. \ref{fig:diffusion_over_height}.
As done for the experiments, they are also scaled to a common porosity of $\varepsilon=0.38$ using the term $\varepsilon^2/(1-\varepsilon)$ from Eq. \ref{eq:dk_derjaguin2}.
The filled symbols are well comparable to the experiments within the experiment uncertainties.

\begin{figure}
  \centering
  \includegraphics[width=\columnwidth]{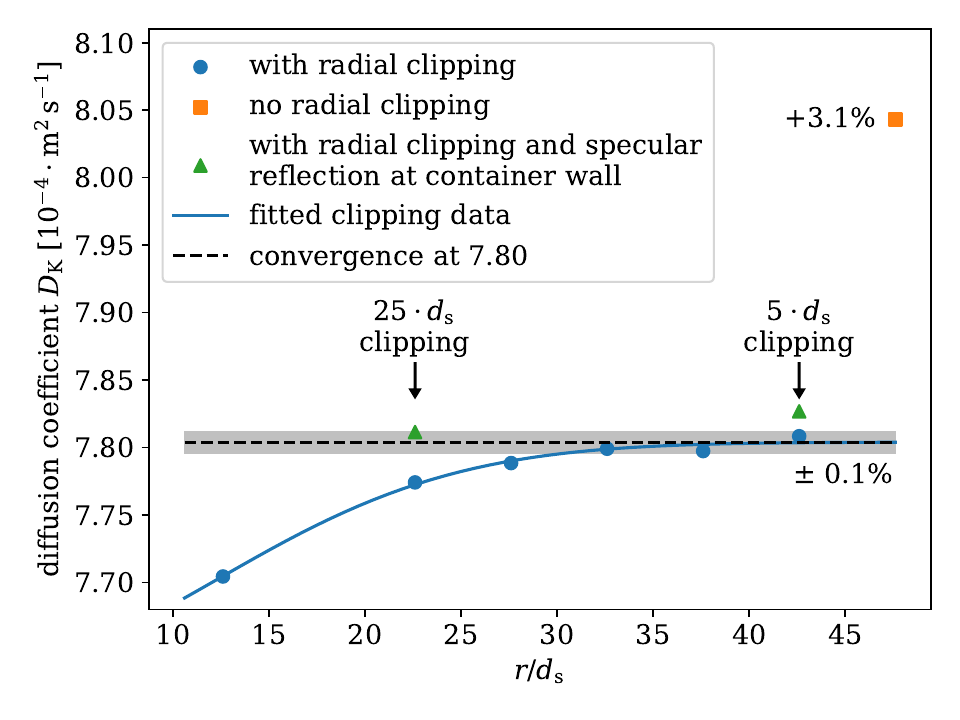}
  \caption{\label{fig:boundary_effect}
  The boundary effect due to boundary porosity (orange) is reduced by radial clipping (blue).
  Clipping of $5 \cdot d_\mathrm{s}$ is our default value.}
\end{figure}

\subsection{Boundary Effects} \label{sect:boundary_effects}

We have seen in Sect. \ref{sect:numerical_samples} that the granular samples show boundary effects in the porosity.
These could not be avoided in the experiments, where boundary conditions naturally occur.
However, we want to study the consequence of the boundary effects and make a prediction for an idealised (infinite) sample.

Averaged over one sphere diameter in radius (Fig. \ref{fig:porosity_waves}, dashed lines), the outer cylinder shell has a higher porosity, thus we expect a higher flow.
To test this, a cylindrical container wall was inserted in the DSMC model at different radii around the centre to clip the outer shells.
In Fig. \ref{fig:boundary_effect}, the orange data point represents the diffusion coefficient of a full (non-clipped) SBD sample with $d_\mathrm{s}=50$ \mum, $d_\mathrm{c}=95.2\cdot d_\mathrm{s}$, and $\varepsilon=0.41$.
The blue circles correspond to different clipping from $5$ to $35 \cdot d_\mathrm{s}$, corresponding to a container diameter $d_\mathrm{c}$ of 1.26 to 4.26 mm, respectively.
The latter still show a boundary effect as the cylinder wall acts as a disturbance (the cylinder boundary is simulated as a surface with diffuse reflection properties, thereby reducing the flow) compared to an infinite sample, which becomes less important for large samples.
We see that the data converges against a stable diffusion coefficient (solid blue and black dashed line) and that a radius of $\sim 30 \cdot d_\mathrm{s}$ is sufficiently large to result in an error in the order of $0.1\%$.

To test the hypothesis of the flow reduction from a diffuse surface, we simulated a sample with clipping $5$ or $25 \cdot d_\mathrm{s}$ and specular reflection at the cylinder wall (green triangles).
The diffusion coefficient is close to the convergence value of the diffuse-reflection case, and a deviation of $0.25\%$ can possibly be attributed to a remaining uncertainty of the convergence value.
This shows that a sample with specular reflection at the boundary walls well represents an infinite sample even for small computation volumes.
Specular reflection was however not utilised in this work beyond this test case.

The effect of the boundary porosity (orange square) amounts to $3.1\%$ if not taken into account.
Whenever we apply radial clipping, we determine the porosity in this cylinder core.

We see similar porosity boundary effects in the vertical direction (Fig. \ref{fig:porosity_waves}, left).
The pressure gradient in the experiments as well as in Sect. \ref{sect:dsmc_experiment} was determined from the pressure difference above and below the sample, which includes boundary effects.
To compensate for these, we do not modify the sample (we use the full height) but we determine the pressure gradient from a fit of the pressure $p(z)$ in a range from $z=5\cdot d_\mathrm{s}$ to $z=h-5 \cdot d_\mathrm{s}$.
This excludes the potentially non-linear entry and exit flow due to the porosity boundary.
Also in this case, we compute the porosity only from the volumes that were used to compute the pressure gradient.

\begin{figure}
  \centering
  \includegraphics[width=\columnwidth]{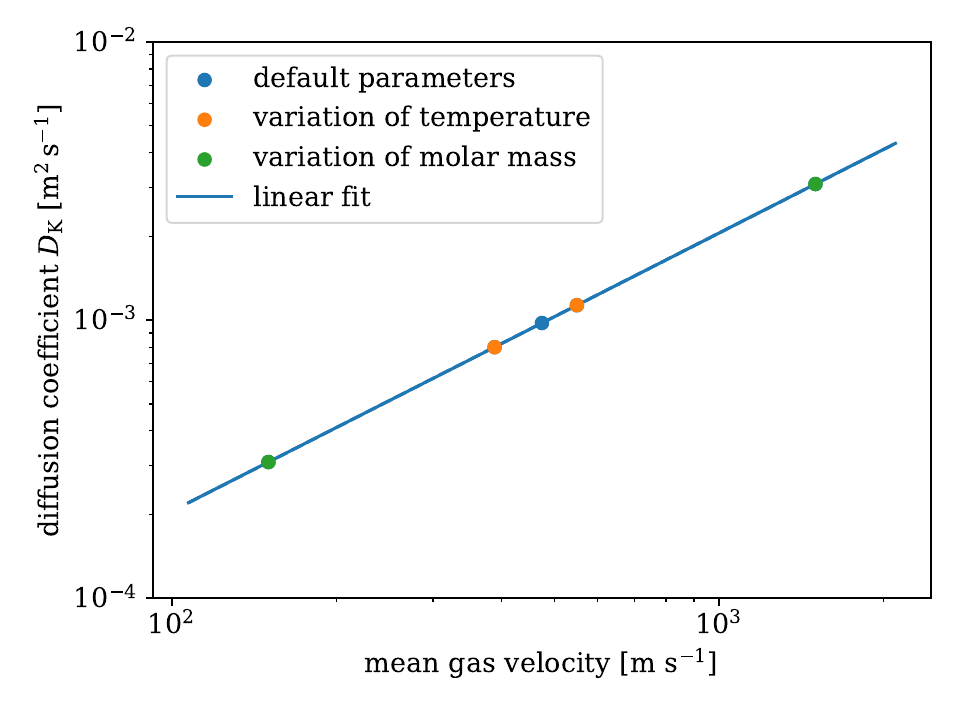}
  \includegraphics[width=\columnwidth]{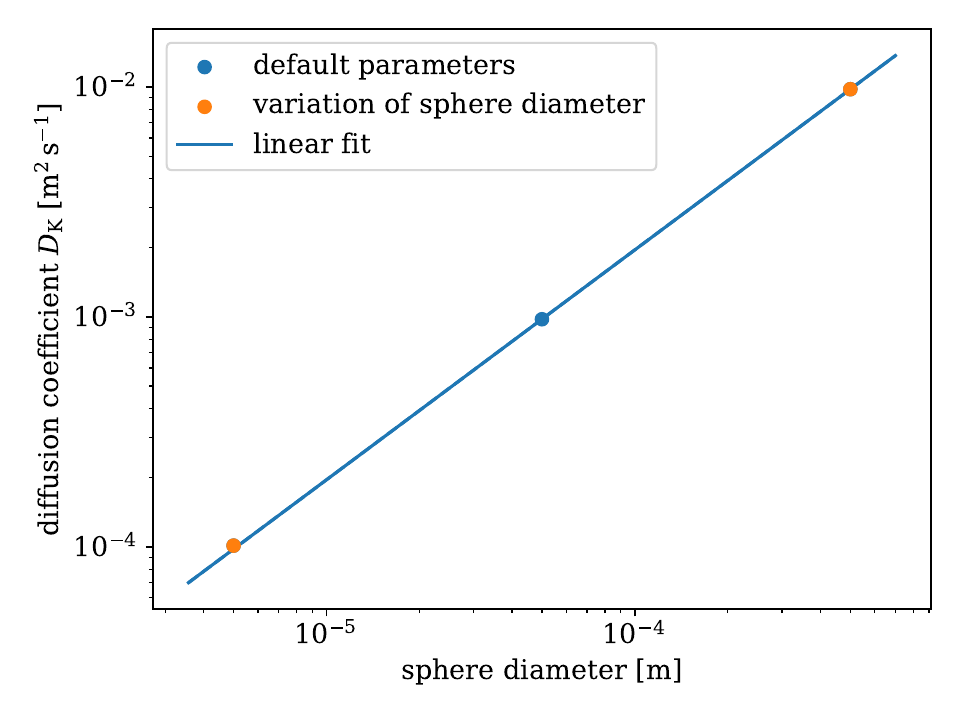}
  \caption{\label{fig:asaeda_velocity_and_diameter}
  Diffusion coefficient from DSMC simulation with variation of the gas velocity (top) and sphere diameter (bottom).
  Both parameters scale linearly (blue lines) as predicted in Eq. \ref{eq:dk_asaeda}}
\end{figure}

\subsection{Scaling of the Diffusion Coefficient} \label{sect:dk_scaling}

According to Eq. \ref{eq:dk_asaeda}, the diffusion coefficient is a function of sample and gas properties.
We tested the  functional dependence on  the sphere diameter $d_\mathrm{s}$, the mean particle velocity $\bar{c}$ (which includes the gas temperature $T$ and the molar mass $M$; Eq. \ref{eq:gas_velocity}), and the porosity $\varepsilon$.
The sample was constructed using SBD with a container of $d_\mathrm{c} = 95.2 \cdot d_\mathrm{s}$ and $h=36 \cdot d_\mathrm{s}$, with defaults \mbox{$d_\mathrm{s}=50$ \mum} and $\varepsilon=0.45$.
We apply a $5 \cdot d_\mathrm{s}$ boundary clipping in all directions as described above.

Figure \ref{fig:asaeda_velocity_and_diameter} shows the variation of the mean gas velocity (top) and sphere diameter (bottom).
The blue symbols refer to the default parameters provided above.
The orange and green symbols show a variation of the temperature (top, orange), molar mass (top, green), and sphere diameter (bottom, orange)
The linear fits (blue solid lines) confirm the linear relations as of Eq. \ref{eq:dk_derjaguin2}.

\begin{figure}
  \centering
  \includegraphics[width=\columnwidth]{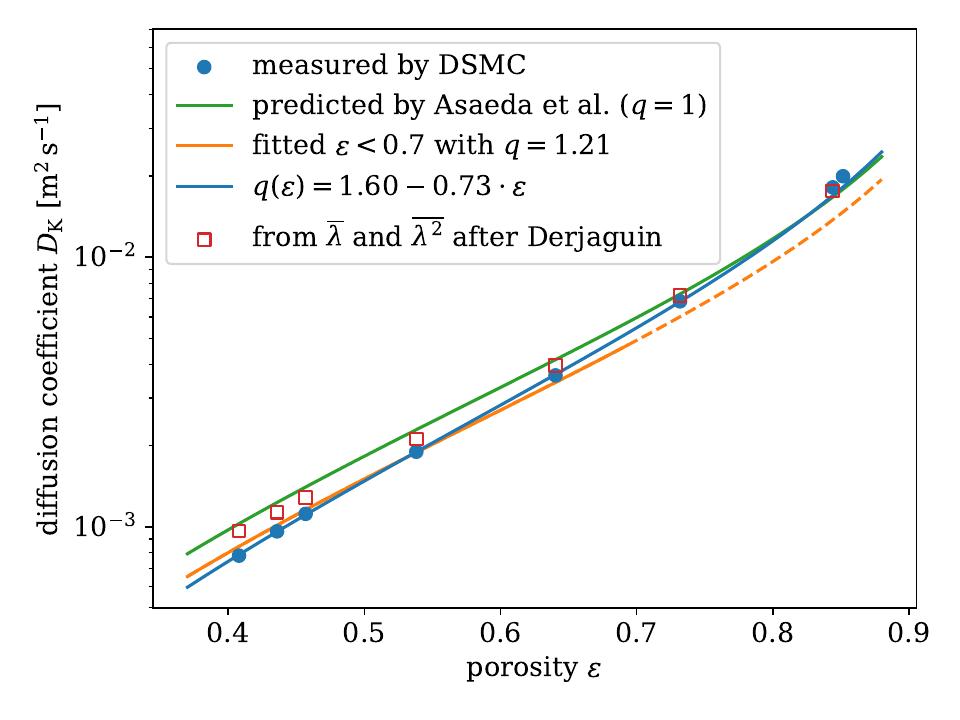}
  \caption{\label{fig:asaeda_porosity}
  Diffusion coefficient from DSMC simulation with variation of the sample porosity (SBD samples).
  Curves corresponds to Eq. \ref{eq:dk_asaeda} with varying $q$, red squares to Eq. \ref{eq:dk_derjaguin1}.}
\end{figure}

Figure \ref{fig:asaeda_porosity} shows the effect of a variation of the sample porosity, with all other parameters being defaults listed above (the value at $\varepsilon=0.45$ is identical to the blue symbols in Fig. \ref{fig:asaeda_velocity_and_diameter}).
The samples were prepared with the SBD method of different $\nu$ parameter.
The green curve corresponds to Eq. \ref{eq:dk_asaeda} with $q=1$ (identical to Eq. \ref{eq:dk_derjaguin2}) and is indistinguishable from a fit to the data with the free parameter $q$.
The orange curve is the same fit but rejecting simulation data of porosities larger than 70\% and resulting in $q=1.21$.
Both show a systematic difference with respect to the data (the data is steeper).

The blue curve is a fit where $q$ is allowed to vary linearly with the porosity, resulting in 
\begin{equation}
    q(\varepsilon) = 1.60 - 0.73 \cdot \varepsilon \; ,  \label{eq:q_eps}
\end{equation}
which matches the data except for the highest porosity ($\varepsilon > 0.8$, which was also not used for fitting).
The reason for the variation of the $q$ value could be in the structural feature of the sample, gradually changing with porosity.
In spite of our intention to use the same algorithm for all porosities, the shape of cavities does still change as a function of porosity.
This could also result in pores, which contribute more or less to the flow and an effective 'flow porosity', but this is hypothetical and needs further study.

The red squares in Fig. \ref{fig:asaeda_porosity} will be described below in Sect. \ref{sect:statistical_analysis}.

\begin{figure*}
  \centering
  \includegraphics[width=\textwidth]{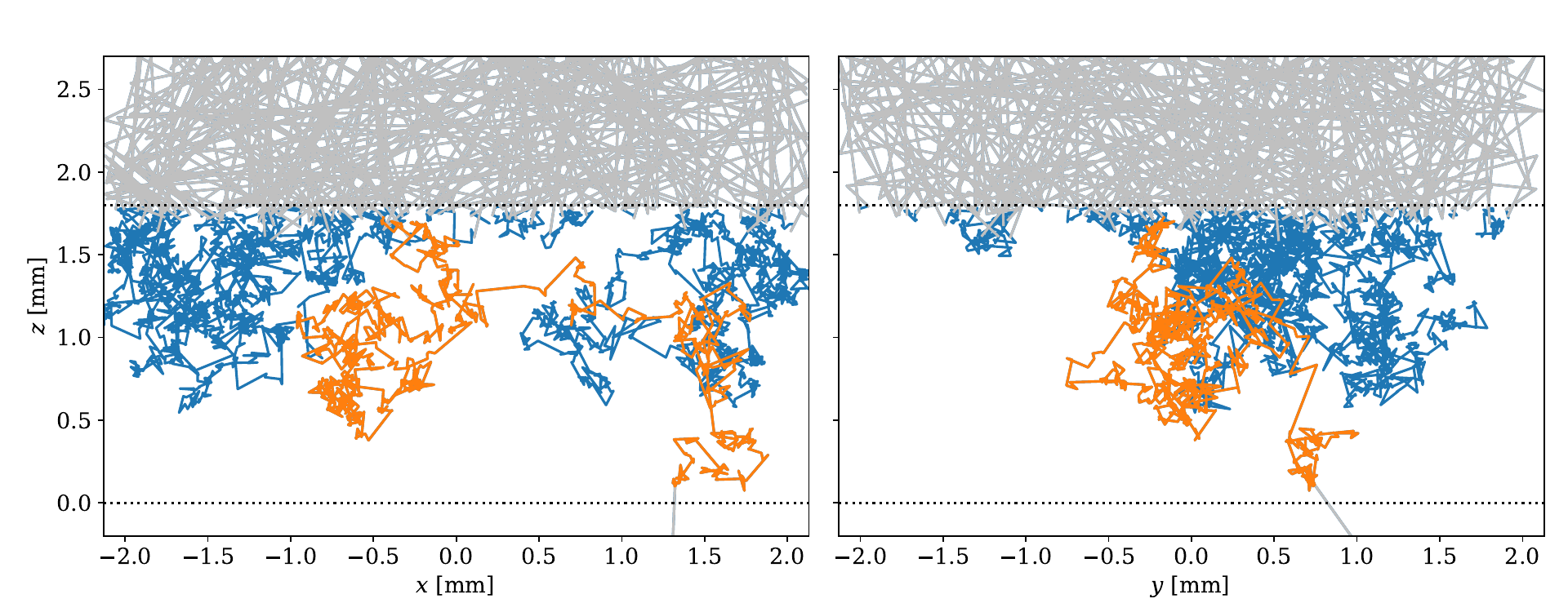}
  \caption{\label{fig:example_particle}
  Example molecule track through a sample (between dotted lines).
  Gray tracks are outside the sample, blue and orange tracks within the sample.
  Blue tracks exit again towards the upstream volume, while the single orange track fully traverses the sample.}
\end{figure*}

\subsection{Statistical Molecule Analysis} \label{sect:statistical_analysis}

\begin{table}
    \centering
    \caption{Results for DSMC simulations with our default parameters as listed in Sect. \ref{sect:dk_scaling} and varying porosity $\varepsilon$ of SBD samples.}
    \label{tab:porosity_series}
    \begin{tabular}{c|c|c|c|c}
        $\varepsilon$ & $D_\mathrm{K}$ [m$^2$ s$^{-1}$] & $\overline{\lambda} / d_\mathrm{s}$ & $\overline{\lambda^2} / \overline{\lambda}^{\;2}$ & $q$ \\
        \hline
        0.41 & $7.8 \cdot 10^{-4}$ & 0.45 & 1.93 & 1.31 \\  
        0.54 & $1.9 \cdot 10^{-3}$ & 0.77 & 1.91 & 1.21 \\  
        0.64 & $3.6 \cdot 10^{-3}$ & 1.16 & 1.96 & 1.14 \\  
        0.73 & $6.9 \cdot 10^{-3}$ & 1.77 & 2.02 & 1.06 \\  
        0.84 & $1.8 \cdot 10^{-2}$ & 3.33 & 2.20 & 0.90 \\  
    \end{tabular}
\end{table}

With DSMC, we have the possibility to perform a statistical analysis of a large number of individual molecules.
The paths of molecules were stored as tracks, listing locations of collision with either walls (spheres or container) or other molecules.
The collisions among molecules amounted to approx. $0.5\%$, so these are negligible (Knudsen regime).
An example for a molecule traversing the sample is shown in Fig. \ref{fig:example_particle}.
It bounces within the upstream volume (top, gray lines) and eventually enters into the sample (track becomes blue).
The motion is not directed, thus it is likely that the molecule returns to the upstream volume and eventually re-enters the sample.
The last track, which finally makes it to the downstream volume (bottom) is marked in orange.
The sample has a porosity of $\varepsilon=0.64$.

\begin{figure}
  \centering
  \includegraphics[width=\columnwidth]{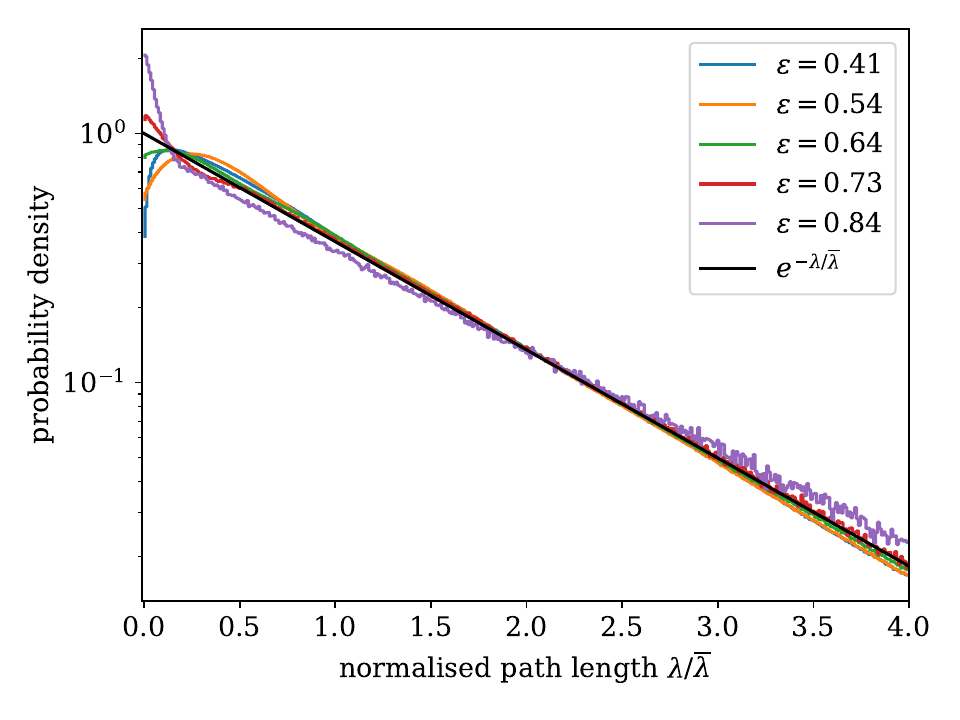}
  \caption{\label{fig:mean_free_paths}
  Probability density distribution of path lengths $\lambda$ for samples of different porosities.
  When normalised with the mean path $\overline{\lambda}$ distributions converge against Eq. \ref{eq:poisson} (black line) for $\lambda \gtrsim \overline{\lambda}$.}
\end{figure}

The length of individual segments between two collisions are the path lengths $\lambda$.
Considering the segments inside the sample (blue and orange in Fig. \ref{fig:example_particle}), Fig. \ref{fig:mean_free_paths} shows a distribution of these paths lengths $\lambda$ for different sample porosities (again we consider the porosity of the centrally clipped cylinder).
To achieve smooth statistics, the number of molecules per simulation was between 2,000 and 12,000, such that the number of path segments was between 9 and 221 million.
When normalising the horizontal axis to the mean path $\overline{\lambda}$, all distributions converge against Eq. \ref{eq:poisson} (for $\lambda \gtrsim \overline{\lambda}$) as postulated by \citet{Derjaguin:1946}.
The shape of the distribution for small $\lambda$ changes significantly and as a function of porosity.
While for low porosities short paths are suppressed (blue, $\varepsilon=0.41$), they are increased for the highest porosity (purple, $\varepsilon=0.84$).
The latter is also the only case where the distribution falls flatter than Eq. \ref{eq:poisson}.
We did not investigate this in greater detail though.

\begin{figure}
  \centering
  \includegraphics[width=\columnwidth]{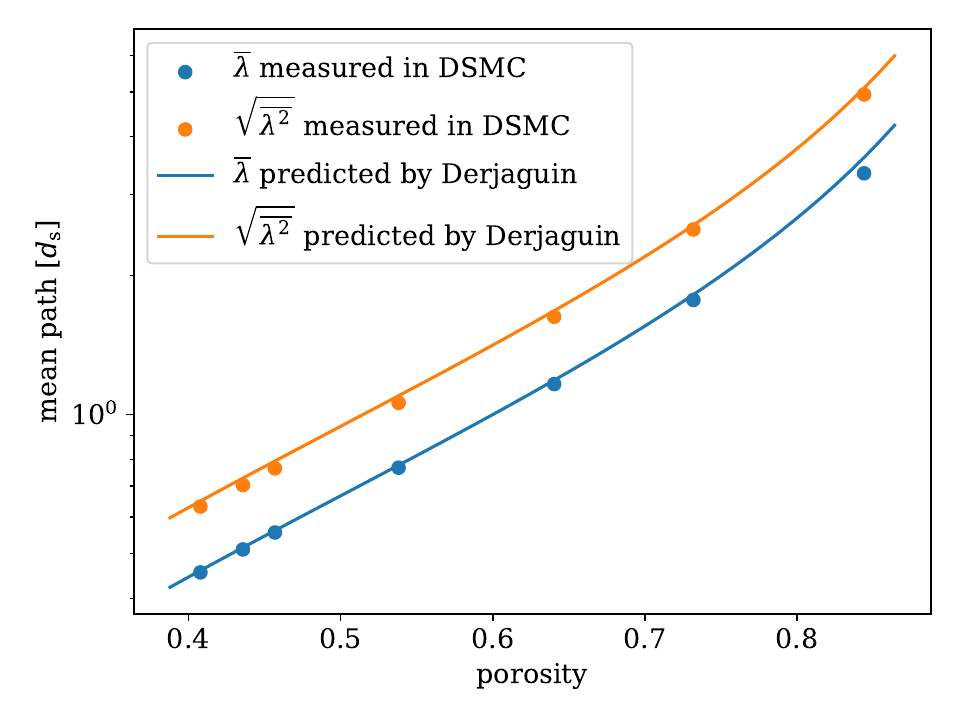}
  \caption{\label{fig:mfp_derjaguin}
  Mean path length and square averaged as measured in DSMC (blue and orange circles) in comparison to the prediction by \citet[][solid lines, Eqs. \ref{eq:lambdas_ratio} and \ref{eq:lambda_derjaguin}]{Derjaguin:1946}.}
\end{figure}

From the path segments, we can directly compute the mean path $\overline{\lambda}$ and the square-averaged path $\overline{\lambda^2}$.
This is presented in Fig. \ref{fig:mfp_derjaguin}, where $\overline{\lambda}$ is displayed as blue and $\overline{\lambda^2}$ as orange circles.
The mean path as predicted by \citet[][Eq. \ref{eq:lambda_derjaguin}]{Derjaguin:1946}, is shown as the blue line.
The match with the DSMC results is excellent, given that the equation has no free parameter and is mostly based on two theoretical assumptions that (a) all volume elements of pores are evenly filled by paths and (b) points of collisions are evenly distributed over the surface of pores.
We conclude that the structure of the sample (i.e., the SBD recipe) cannot have a large influence on the mean paths as \citet{Derjaguin:1946} did not make any assumptions on this except that it can be described with a single porosity (i.e., a constraint on homogeneity).
Applying Eq. \ref{eq:lambdas_ratio}, we can plot the orange line in comparison to the $\overline{\lambda^2}$ values computed directly from DSMC results.
The match is not a surprise given that the distributions in Fig. \ref{fig:mean_free_paths} are largely described by Eq. \ref{eq:poisson}.
The values are presented in Table \ref{tab:porosity_series}, the ratios are matching within $10\%$.

From the mean paths ($\overline{\lambda}$ and $\overline{\lambda^2}$), we can compute the diffusion coefficient based on the statistical model of \citet{Derjaguin:1946} after Eq. \ref{eq:dk_derjaguin1}.
This is displayed in Fig. \ref{fig:asaeda_porosity} as red squares.
These differ with respect to the coefficients determined from the overall pressure gradient (blue symbols) up to $15\%$ (smallest porosity).
Although less prominent than for the blue symbols, also these values are steeper than the \citeauthor{AsaedaEtal:1973} model (Eq. \ref{eq:dk_asaeda}), which is obtained from \citeauthor{Derjaguin:1946} Eq. \ref{eq:dk_derjaguin1} with the simplification of Eq. \ref{eq:lambdas_ratio}.

\begin{figure}
  \centering
  \includegraphics[width=\columnwidth]{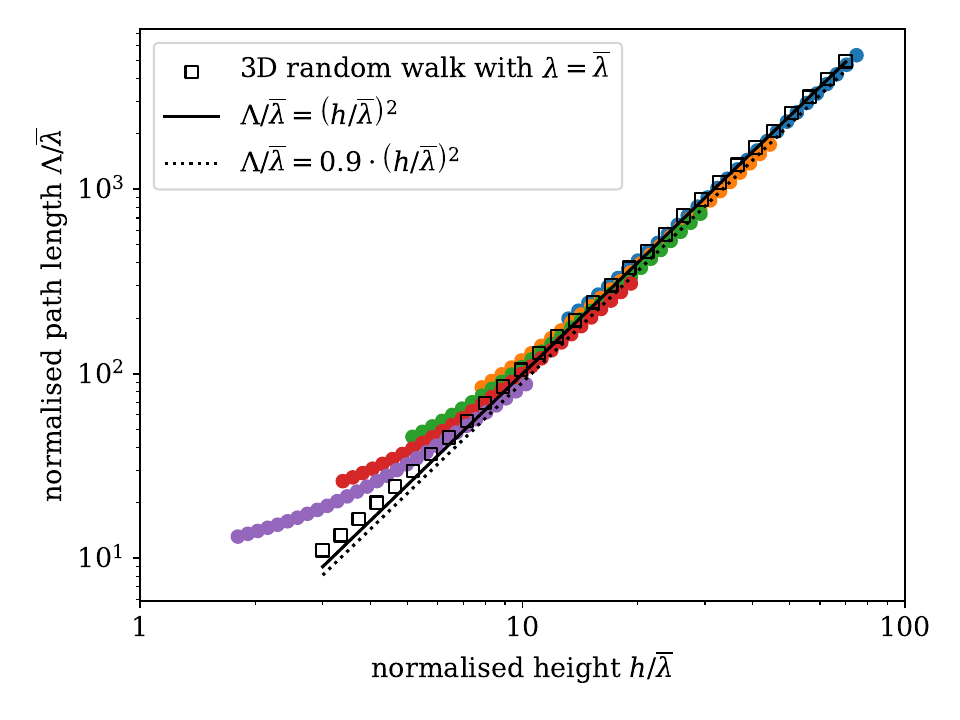}
  \caption{\label{fig:diffusion_square}
  When normalising to the mean path, the total travelled distance converges against the square of the fill height for all porosities.
  Coloring is identical as in Fig. \ref{fig:mean_free_paths}.}
\end{figure}

We can show that the process is described by an idealised diffusion.
Figure \ref{fig:diffusion_square} shows the total travelled distance $\Lambda$ (the sum of all segments of length $\lambda$) as a function of the reached height.
We split the sample into 30 fractions of the fill height and determine which distance $\Lambda$ a molecule travelled to reach this height for the first time.
To increase the statistics, we consider each entry of a molecule into the sample as an individual molecule.
For each height we then average the travelled distance over all molecules that reach this height.
The different porosities in different colors all converge against a line of
\begin{equation}
    \frac{\overline{\Lambda}}{\overline{\lambda}} = \alpha \cdot \left(\frac{h}{\overline{\lambda}}\right)^2  \label{eq:ficks_law}
\end{equation}
with $\alpha \approx 0.9$ (dotted line), which is close to the idealised diffusion with $\alpha=1$.
The latter is shown as black squares, representing a random walk of an undisturbed particle, where the direction of motion in 3D is randomised after each distance $\lambda = \overline{\lambda} = 1$.
For simplicity, we take the reached distance as the radial distance from the start point.
As expected, this converges against the solid black line with $\alpha=1$.
Also this simplified case shows a deviation for small distances and converges towards Eq. \ref{eq:ficks_law}.

\section{Discussion of Results} \label{sect:discussion}

\begin{figure}
  \centering
  \includegraphics[width=\columnwidth]{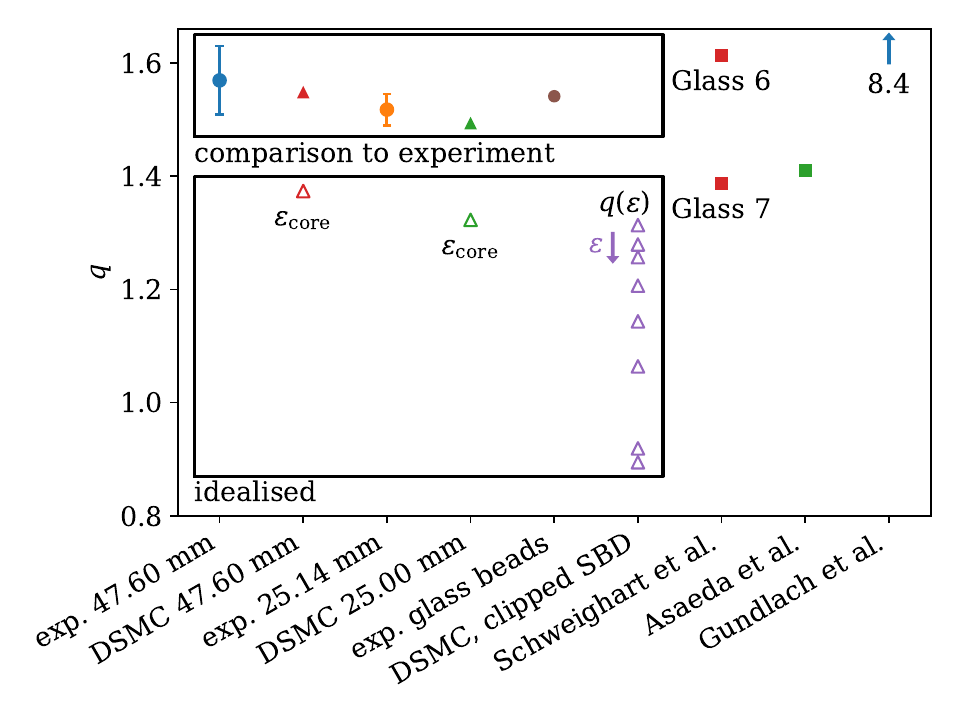}
  \caption{\label{fig:dk_comparison}
  Comparison of diffusion coefficients ($q$ from Eq. \ref{eq:dk_asaeda}) from our work and the literature.}
\end{figure}

The results of the diffusion coefficient are compiled in Fig. \ref{fig:dk_comparison}, where our simulations and experiments are framed by the black boxes to the left.
We focus on the upper box first.
To avoid scaling of different sphere diameters and porosities, the data are presented as the $q$ parameter introduced by \citet[][our Eq. \ref{eq:dk_asaeda}]{AsaedaEtal:1973}.
Our experiments with 0.5 mm steel beads (blue and orange) apply the standard deviation as error bars.
As already seen in Fig. \ref{fig:diffusion_over_height}, the DSMC simulations (filled triangles) match our experiments within the experimental uncertainties.
The experiment with glass beads (brown circle) was performed only for one filling, and fluctuations between the three repeats are within the symbol size.

Experiments by \citet{SchweighartEtal:2021} are displayed as red squares.
Their experiment ``Glass 6'' with an average sphere diameter of 0.43 mm and porosity 0.37 requires the smallest scaling with respect to our parameters and matches well.
Their experiment labelled ``Glass 7'', where the spheres are identical to our glass experiments (brown circle) matches less well with our results.
We have seen in our work that the results are extremely sensitive on porosity, so it is noteworthy that the porosity for the two is very different (also see their Fig. 13), which may explain the variation.

The experiments by \citet{AsaedaEtal:1973} are summarised by a value of $q=1.41$ (green square) over a wide range of sphere diameter (5.7 to 870 \mum) and porosities (0.38 to 0.45).
The theoretical expectation from \citet{Derjaguin:1946} is defined as $q=1$, where Eqs. \ref{eq:dk_derjaguin2} and \ref{eq:dk_asaeda} become identical.

Results by \citet{GundlachEtal:2011b} deviate by a factor $\sim 5$, which is not understood.
An effort was taken with the authors of that paper to understand this discrepancy, including a repeated measurement of the particle sizes and the impact of the supporting filter paper, but it could not be explained.
Although purely hypothetical, the issue might be a miscalibration of the pressure sensor but this cannot be reproduced.

In DSMC we had the chance to reduce boundary effects to a minimum by considering porosity and diffusion only in the homogeneous core of the sample.
This is the closest we can go to an idealised, infinite sample, and is represented as purple symbols (data in Table \ref{tab:porosity_series}).
As we have seen in Eq. \ref{eq:q_eps}, the $q$ values fall with increasing porosity, thus the uppermost purple triangle corresponds to $\varepsilon=0.41$ and the lowermost to $\varepsilon=0.73$.
We have an uncertainty for the pure RBD sample ($\varepsilon=0.85$), which deviates from Eq. \ref{eq:q_eps} ($q=0.9$) and also deviates in the distribution of mean paths (Fig. \ref{fig:mean_free_paths}).
This outcome is plausible since for extremely high porosity the random packing creates branch-like structures of spheres, leaving an anisotropic pore structure with longer void dimensions along the net flow direction.
As a consequence, very long path lengths appear with a slight predominance of directions parallel to the net flow.
Although they occur to a much lesser relative amount as with the smaller porosities, they can increase the flux above the corresponding flux of an analogous sample with exactly isotropic pore structure.

The results of the experiments can be compared to the idealised DSMC results, when we take into account the boundary conditions.
The red and green open triangles correspond to the simulations that reproduce the experiments, including porosity boundary effects, but as a porosity we use the core porosity of a central cylinder clipped by $5 \cdot d_\mathrm{s}$ in all directions (described in Sect. \ref{sect:boundary_effects}).
The $q$ values become consistent with the idealised SBD sample of $\varepsilon = 0.41$ and $q=1.31$ (uppermost purple triangle).
The rationale to use the core porosity is that the overall sample volume is dominated by this porosity, which determines the diffusion.
There are still a few minor differences in the setup (different bulk porosities, vertical porosity gradient in LIGGGHTS sample, uncorrected boundary flux), but this major correction establishes a satisfactory link between experiments and idealised samples.

We conclude that the diffusion is well described by Eq. \ref{eq:dk_asaeda} \citep{AsaedaEtal:1973, Derjaguin:1946}, with a refinement for the structure (or porosity) that we describe as $q(\varepsilon)$ according to \mbox{Eq. \ref{eq:q_eps}}.
The microphysical reason for this $q(\varepsilon)$ factor remains to be studied.
Boundary effects from the reduced porosity at container walls affect the experimental measurements and most probably also those by \citet{SchweighartEtal:2021} and \citet{AsaedaEtal:1973}.
Additional boundary effects influencing the experimental measurements are described by \citet{LaddhaEtal:2023}.
Most of these are avoided in the presented work, by paying special attention to the measurement setup and by a careful sample selection and preparation.

\section{Conclusion} \label{sect:conclusion}

We find a good match between the determined diffusion coefficient in experiments and our DSMC simulations.
The DSMC method is moreover ideal to study diffusion in a granular bed, providing insights into macroscopic parameters (diffusion coefficient) as well as a microscopic description (mean path, path-length distribution).

The model of \citet{Derjaguin:1946} is excellent in predicting the scaling of the diffusion coefficient with gas and sample properties and quite reasonable in predicting its absolute values.
An additional factor $q$ introduced by \citet{AsaedaEtal:1973} was found to be depending on the structure or porosity of the samples and we provide a refinement to Eq. \ref{eq:dk_asaeda}, applicable to packing of spheres, as $q(\varepsilon)$ provided in Eq. \ref{eq:q_eps}.
This factor is a kind of tortuosity, but not equal to the tortuosity applied in models based on a representation of the pore space as a collection of cylindrical filaments, where the approach by \citet{Knudsen:1909} is used.

On a microscopic level, knowledge of the mean path length $\overline{\lambda}$ and square-average path length $\overline{\lambda^2}$ can be used to determine the diffusion coefficient from Eq. \ref{eq:dk_derjaguin1} as predicted by \citet{Derjaguin:1946}.
His model, applied to a geometry of packed beds of spheres, moreover allows to predict the mean path according to Eq. \ref{eq:lambda_derjaguin}.
The distribution of mean paths converges against Eq. \ref{eq:poisson} but deviates for small paths, depending on porosity.
This results in an uncertainty of the squared average path (Eq. \ref{eq:lambdas_ratio}) of $10\%$, propagating into Eq. \ref{eq:dk_derjaguin2}.

Applied to thermophysical models of cometary surfaces and a structure model of porous pebbles \citep{BlumEtal:2014}, the equation system provided above (and verified by experiments and numerical simulations) provides all necessary information to compute the diffusion coefficient inside or between pebbles.
This works for idealised systems, where pebbles can be described by spheres and the pore space is empty (i.e., not filled with fractals as proposed by \citet{FulleBlum:2017}).
We plan to continue this work to make it applicable to more complex samples.
In particular the structure used by \citet{FulleEtal:2020}, required to explain the disintegration of pebbles, will be an interesting study case.
Based on Rosetta dust observations \citep{GuettlerEtal:2019}, \citet{CiarnielloEtal:2023} pointed out that this requires size ratios between pebbles and constituent grains in the order of $10^5$ (their factor $\chi$).
This poses an interesting challenge on the design of the DSMC model, performing computations on very different scales.

\section*{Acknowledgment}
We thank M. Strowitzki and F. Giessmann for their support in the development of the vacuum system.

\section*{Data Availability}
The data underlying this article will be shared on reasonable request to the corresponding author.

\bibliographystyle{mnras}
\bibliography{literature}
\label{lastpage}
\end{document}